# Tuning across the BCS-BEC crossover in the multiband superconductor $Fe_{1+y}Se_xTe_{1-x}$ : An angle-resolved photoemission study

S. Rinott[1], K.B. Chashka[1] A. Ribak[1], E. D. L. Rienks[2,3] A. Taleb-Ibrahimi[4], P. Le Fevre[4], F. Bertran[4], M. Randeria[5] and A.Kanigel[1*]

[1] Department of Physics, Technion - Israel Institute of Technology, 32000 Haifa, Israel,

[2] Leibniz-Institut für Festkörper- und Werkstoffforschung, Dresden, D-01171 Dresden, Germany,

[3] Institut für Festkörperphysik, Technische Universität Dresden, D-01062 Dresden, Germany

[4] Synchrotron-SOLEIL, Saint-Aubin, BP 48, F91192 Gif sur Yvette Cedex, France

[5] Department of Physics, Ohio State University, Columbus, OH 43210, USA

*Corresponding author. Email: amitk@physics.technion.ac.il

The crossover from Bardeen-Cooper-Schrieffer (BCS) superconductivity to Bose-Einstein condensation (BEC) is difficult to realize in quantum materials because, unlike in ultracold atoms, one cannot tune the pairing interaction. We realize the BCS-BEC crossover in a nearly compensated semimetal $Fe_{1+y}Se_xTe_{1-x}$ by tuning the Fermi energy $\varepsilon_F$ via chemical doping, which permits us to systematically change $\Delta/\varepsilon_F$ from 0.16 to 0.50, where $\Delta$ is the superconducting (SC) gap. We use angle-resolved photoemission spectroscopy to measure the Fermi energy, the SC gap and characteristic changes in the SC state electronic dispersion as the system evolves from a BCS to a BEC regime. Our results raise important questions about the crossover in multiband superconductors which go beyond those addressed in the context of cold atoms.

The Bardeen-Cooper-Schrieffer (BCS) to Bose-Einstein condensation (BEC) crossover [1] has emerged as a new paradigm for pairing and superfluidity in strongly interacting Fermi systems, which interpolates between two well-known limiting cases: BCS theory for fermions with a weak attractive interaction and a BEC of tightly-bound bosonic pairs. This crossover has now been extensively investigated in ultracold Fermi gas experiments [2 - 4], where the strength of the attraction between Fermi atoms is tuned using a Feshbach resonance. This has led to a number of new insights [5 - 7] into the very strongly interacting unitary regime that lies in between the BCS and BEC limits. One of the key characteristics of the crossover regime is that the ratio of the pairing gap $\Delta$ to the bandwidth or Fermi energy $\varepsilon_F$ is of order unity, as opposed to the BCS limit where $\Delta/\varepsilon_F \ll 1$.

From this perspective, most superconducting (SC) materials, (even strongly correlated systems), such as the cuprates, are much closer to the BCS limit than in the crossover regime. A key obstacle to realizing the BCS-BEC crossover in quantum materials is that one does not have the ability to continuously tune the strength of the pairing interaction that controls $\Delta$. We offer an alternative path to the realization of the BCS-BEC crossover: Instead of trying to change $\Delta$, we tune $\varepsilon_F$ by chemically doping carriers and drive the system through the BCS-BEC crossover.

We focus on the Fe-based superconductor $Fe_{1+y}Se_xTe_{1-x}$, which is a nearly compensated semimetal [8 - 10], and tune the Fermi energies of various pockets by changing the excess Fe concentration $y$. Using angle-resolved photoemission spectroscopy (ARPES), we systematically map out the evolution of the electronic excitation spectrum to demonstrate that we traverse the BCS-BEC crossover in the solid state. Our work raises fundamentally new questions about the crossover in a multiband superconductor, which go beyond the single band physics explored in cold atoms.

**Results**

Our main results are summarized in Fig. 1, where we show the ARPES intensity at 1K, well below SC $T_c$, for a hole band near the $\Gamma$ point of the Brillouin zone. The top three panels (Fig. 1 A to C) correspond to three samples in order of decreasing $y$ the amount of excess Fe. We deduce the following important results from these data.

(1) From the electronic dispersion far from the chemical potential, which is the same in the normal and SC states, we estimate $\varepsilon_F$, the unoccupied bandwidth for this hole pocket, and show that it systematically decreases with decreasing $y$.

(2) We measure the SC energy gap $\Delta$ from the energy of the coherence peaks in the spectra, which correspond to the Bogoliubov quasiparticles. We find $\Delta = 3 \pm 0.5$ meV for all three samples. We conclude that the dimensionless measure of the pairing strength $\Delta/\varepsilon_F$ =0.16, 0.3 and 0.5 increases monotonically with decreasing $y$, exhibiting a crossover from a BCS to a BEC regime.

(3) As $\Delta/\varepsilon_F$ increases, we see that the dispersion of the coherence peak changes from being BCS-like dispersion with characteristic back-bending near $k_F$, to the unusual BEC-like dispersion, with a minimum gap at $\mathbf{k}=0$. This can also be viewed as the shrinking of the "minimum gap locus" [11] from a contour enclosing a finite area in the large $y$ BCS regime to a single point in the small $y$ BEC regime, consistent with the prediction [12] of the crossover theory for multiband SCs [13,14].

We show in Fig. 1 (D to F) the spectral function obtained from a simple model and describe below how it captures important features of the BCS-BEC crossover.

Our work builds on recent experimental progress in Fe(Se,Te). We have previously shown [15] that a small $y$ sample had a tiny $\varepsilon_F$ of a few millielectron volts, and a large $\Delta/\varepsilon_F$ that places it well outside the weak coupling BCS limit. The small value of $\varepsilon_F$ in $Fe_{1+y}Se_xTe_{1-x}$ has also been reported by Okazaki et al. [16] which focused on an unoccupied electron band just above $\varepsilon_F$. Small values of $\varepsilon_F$ were also found in FeSe [17], together with the observation [18] of anomalously large SC fluctuation effects attributed to preformed pairs above $T_c$. The results presented here go beyond these earlier works in that we show how a systematic change in $\varepsilon_F$ via doping permits us to tune across the BCS-BEC crossover.

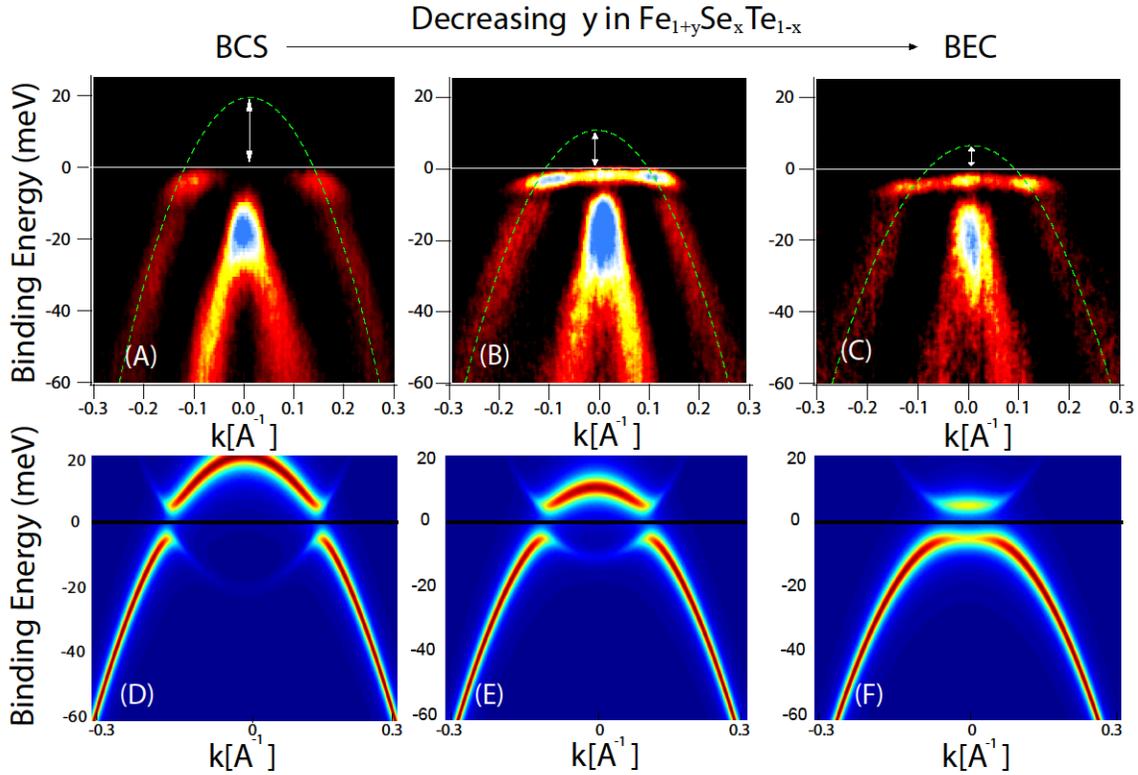

Fig. 1 **SC state ARPES spectra** (A,B,C): ARPES spectra for three samples $Fe_{1+y}Se_xTe_{1-x}$ in order of decreasing *y* (excess Fe) from left to right. The spectra are normalized using the intensity from high-order photons and a constant background is removed. The spectra are sharpened by adding to the original data a small part of its second derivative. The green dashed line is the best fit to the data using a simple parabolic dispersion. (D,E,F): Spectral functions, calculated using the model and parameters described in the text, to describe the BCS to BEC crossover seen in the data in top panels.

**Materials and transport data**

Bulk FeSe has been intensively investigated. It exhibits nematic ordering without antiferromagnetic long-range order (in contrast to other Fe-pnictides) [19] and becomes a $T_c = 65K$ superconductor in monolayer thin-film form on various substrates [20 - 22]. Here, we focus on the lower $T_c$ bulk material Fe(Se,Te), which is nevertheless the most strongly correlated in this family with the largest $\Delta/\varepsilon_F$ ratios, as we show below.

A series of $Fe_{1+y}Se_xTe_{1-x}$ samples were prepared using the modified Bridgman method. We chose to fix x=0.4 near the maximum $T_c \simeq 15K$ for y near zero. The composition x was measured by energy-dispersive x-ray (EDX) analysis. Recently it was shown that by annealing in oxygen it is possible to reduce y, the amount of excess Fe [23 - 27]. Attempts, made by us and other groups, to quantify y using standard techniques, such as x-ray diffraction, and inductively coupled plasma, resulted in inconclusive results. It seems that the only reliable way to measure y is to simply count the number of excess Fe atoms using an STM [27]. We note that a quantitative determination of y in our samples is not essential to our study, since we will focus on the resulting systematic changes in $\varepsilon_F$ measured by ARPES, as shown below.

The result of reducing *y* is an increase in Tc and a change in the resistivity as a function of the temperature. In addition, it was found that the annealing process changes the sign of the Hall resistivity at low temperatures. The samples shown in Fig. 1 are ordered from large value of *y* (BCS) to small value of y (BEC) based on their $T_c$ and resistivity curves.

In Fig. 2 (B and C) we show dc transport data for two samples: The small *y* sample (red curve) has an SC $T_c = 14K$, whereas the large *y* sample (blue curve) has $T_c = 12K$. We verified using a SQUID that all the crystals used have a full SC shielding volume fraction. Note that the entire T-dependence of the resistivity in Fig. 2 (B) is different for the two samples.

The Hall resistance $R_H$ for the same samples shown in Fig. 2 (C) also depends strongly on $y$. At low temperatures, the small $y$ sample has a negative $R_H$, whereas the large $y$ sample has a positive $R_H$, in agreement with previous reports [23]. A quantitative analysis of the Hall data is difficult in view of the multiple bands involved.

**Electronic structure**

To understand our ARPES results, it is useful to first look at the schematic of the band structure of Fe(Se,Te) see Fig. 2 (A).

In bulk FeSe, $Fe^{2+}$ is in a $3d^6$ configuration. With an even number of electrons per unit cell, FeSe is a compensated semimetal, which should be unaffected by the isovalent substitution of Se by Te. Changing $y$ in $Fe_{1+y}Se_xTe_{1-x}$ alters the balance between electrons and holes in what was an almost perfectly compensated semimetal [23 - 25, 28].

In Fig. 2 (A) we show the three hole bands $\alpha_i$ (i=1,2,3 in the notation of Tamai et al. [29]) near the $\Gamma$ point and two electron bands (due to zone folding) near the M point of the Brillouin zone. The red curves in Fig. 2(A) denote the bands for small $y$ sample, whereas the blue curves correspond to large $y$ sample. The main focus of our attention in this paper is the light hole band $\alpha_2$, but we will also discuss some features of the heavy hole band $\alpha_3$ (that remains at or just below the chemical potential) and the electron band near $M$. The $\alpha_1$ band is always well below the chemical potential. A key observation below is that changing excess Fe does *not* just shift the chemical potential in a rigid band structure. Some bands shift in energy, whereas others do not. In addition changing $y$ leads to mass renormalizations.

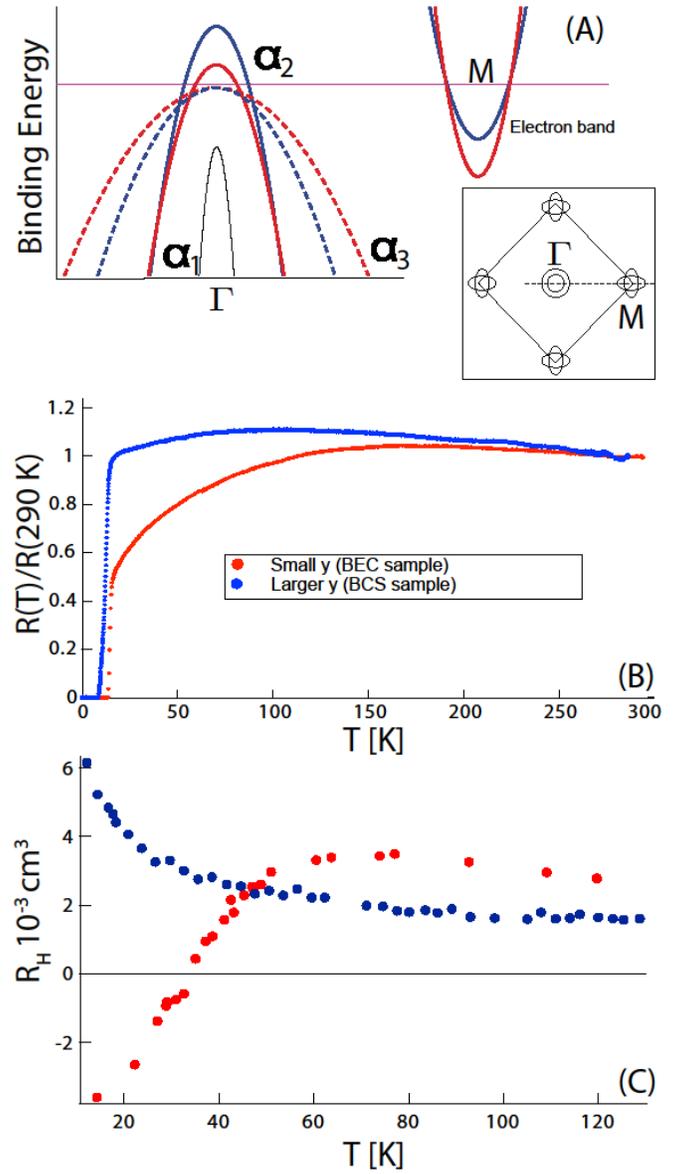

**Fig. 2 Schematic Electronic structure and transport data for $Fe_{1+y}Se_xTe_{1-x}$ samples with different amounts of excess Fe $y$.** In all panels, the red (blue) curve correspond to the small (large) $y$ sample. (A): Schematic band structure and the effect of $y$ on various bands; see text for details. (B): The resistivity as function of temperature. The small $y$ sample has a superconducting $T_c = 14$ K while large $y$ sample has $T_c = 12$ K. (C): Hall resistance $R_H$ of the same two samples. $R_H$ of the small $y$ sample changes sign at around 40K, while that for the large $y$ sample is always positive.

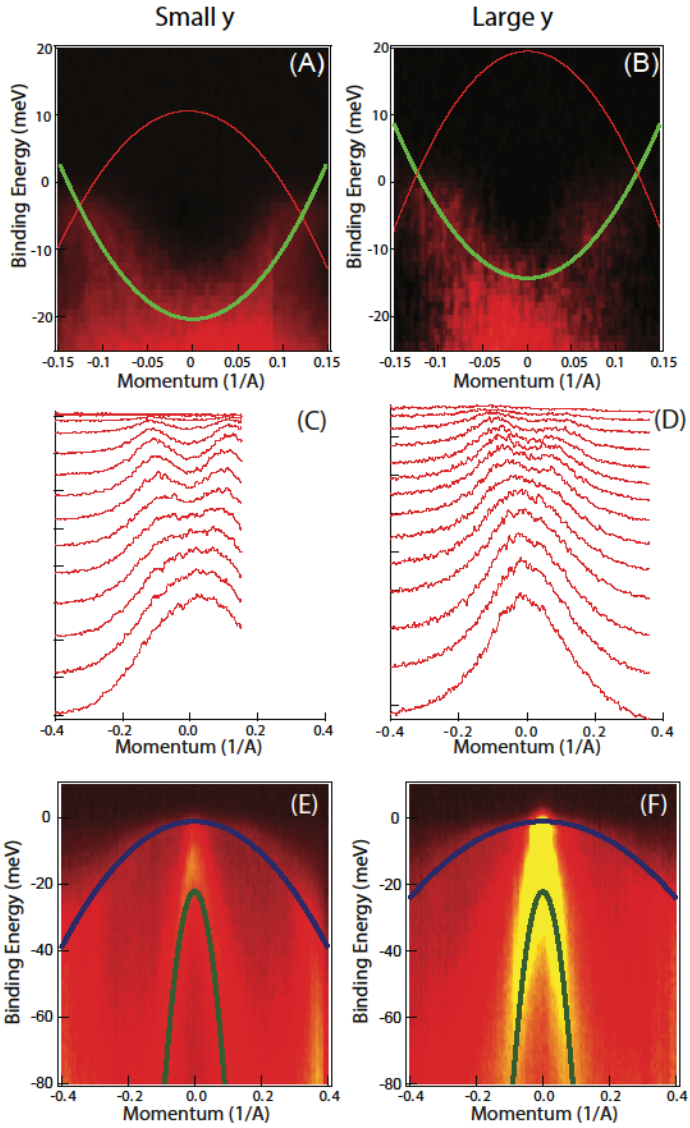

Fig. 3 **ARPES data showing the effect of y on the band-structure.** (A,B): ARPES spectra around the $M$ point for two samples with small and large amount of excess Fe, respectively. A shallow electron pocket can be seen, whose occupied bandwidth decreases with excess Fe. The green lines are best fits using a simple parabolic model to the MDCs shown in panels (C) and (D). The red lines represent the $\alpha_2$ dispersion for the same samples.
(E,F): ARPES spectra of the same two samples around the $\Gamma$ point using vertically polarized 22eV light. In this polarization $\alpha_1$ and $\alpha_3$ can be seen.
The blue curves represent the dispersion of $\alpha_3$ and the green lines that of $\alpha_1$.

**ARPES data for electron and hole bands**

In Fig. 3 (A to D) we compare ARPES data for the electron pocket around the $M$ point for the two samples in Fig. 2. We show the ARPES intensity in Fig. 2 (A) and the momentum distribution curves (MDCs) in Fig. 2 (C) for the small $y$ sample, while Fig. 2 (B and D) show the corresponding data for the large $y$ sample. The green lines in Fig. 2 (A and B) represent the best parabolic fits to the MDC maxima shown in (C and D, respectively). We see that the Fermi energy (occupied bandwidth for the electron band) changed from 20 meV (small $y$ sample) to 14 meV (large $y$ sample), whereas the effective mass is 3.7 $m_e$ for both. We can compare the change in the electron pocket dispersion (green lines in Fig. 2 (A) and (B)) with excess Fe content $y$, to the corresponding $\alpha_2$ dispersion (shown as red curves). The latter is obtained from the measured $\alpha_2$ MDC peaks around the $\Gamma$ point; see Fig. 4. With increasing $y$, as the electron dispersion moves up in energy near the $M$ point, so does the hole dispersion near $\Gamma$. This is qualitatively consistent with the increasing hole-like character of the Hall data in Fig. 2, with increasing $y$. We note that in the ARPES geometry we use (vertical polarization along the $\Gamma - M$ direction) we observe only one of the electrons pockets and we do not observe any coherence peaks in the SC state spectra.

In contrast to the electron and $\alpha_2$ hole bands, we do not find any significant shift in $\alpha_1$ and $\alpha_3$ bands as the excess Fe is changed. In Fig. 3 (E and F) we show the ARPES intensities for the small $y$ (panel (E)) and large $y$ (panel (F)) samples, measured along a cut going through the $\Gamma$-point, using a vertically polarized light. In this polarization, only the $\alpha_1$ and $\alpha_3$ bands can be seen (and not $\alpha_2$). For all the samples that we have measured, the top of the $\alpha_1$ band is at $22 \pm 2$ meV, i.e., deep on the occupied side, in agreement with refs. [15, 29, 30].

We find that the $\alpha_3$ hole band always has a rather low spectral weight. Nevertheless, we see that the unoccupied bandwidth is very small ($\varepsilon_F = -1 \pm 2$ meV) and does not change with $y$. Whether $\alpha_3$ creates a hole pocket or not is not clear [28, 30], although in FeSe $\alpha_3$ is clearly below the chemical potential [31].

The effective mass does change with excess Fe, from 26 $m_e$ for the small $y$ sample to 16 $m_e$ for the large $y$. Effective mass changes of the $\alpha_3$ band were reported also in studies of the effect of Se-Te substitution on the band structure in $Fe_{1+y}Se_xTe_{1-x}$ [32].

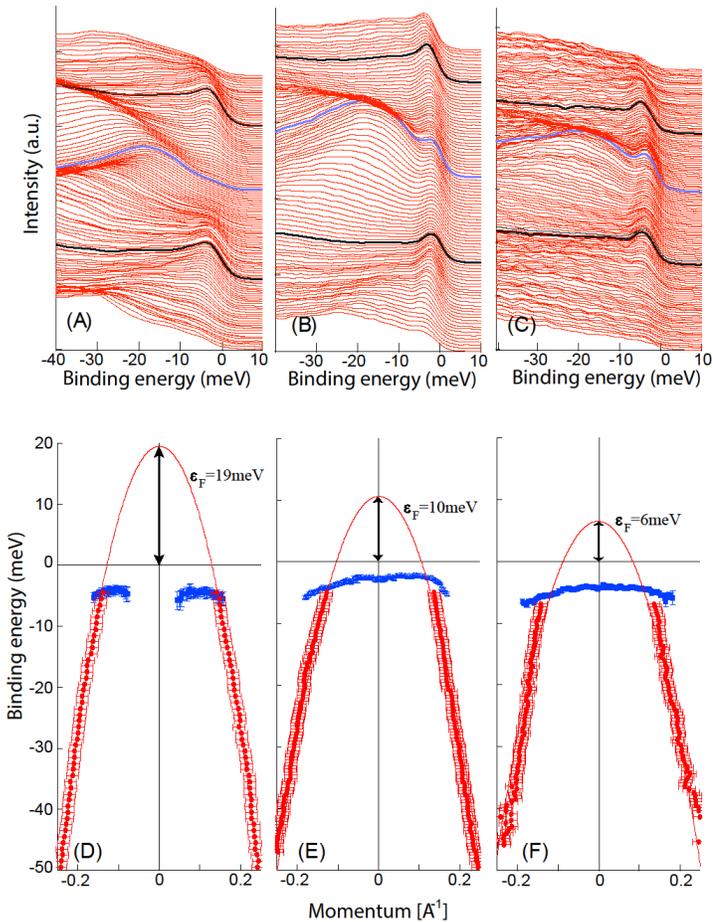

**Fig. 4 Coherence peaks dispersion in the SC state**
(A,B,C): EDC's of same three samples shown in Fig. 1(A,B,C), measured at 1K using horizontally polarized 22eV photons. The blue line the EDC at the $\Gamma$ point and the black lines are the EDC at $k_F$
(D,E,F): The blue dots represent the dispersion of the EDC coherence peaks extracted from panels (A,B,C) respectively. The red dots represent MDC peak positions for binding energies between -5 and -50 meV. The red lines are fits to the latter using a simple parabolic model.

**ARPES evidence for BCS-BEC crossover**
We finally describe the central results of this paper for the $\alpha_2$ hole band (highlighted in Fig. 1).

We show in Fig. 4 (A to C) the energy distribution curves (EDCs) for the three samples in Fig. 1 (A to C) respectively. The large $y$ sample in panel Gif. 4 (A) and the small $y$ sample in Fig. 4 (B) are the same ones for which we presented transport data above. The sample in panel (C) has the smallest value of $y$, the amount of excess Fe and a $T_c = 14.5$ K. The data in Figs. 1 and 4 were obtained using linearly polarized 22eV photons at a temperature of 1K (well below the SC $T_c$'s of the samples) using a horizontal light polarization [33], in which only the $\alpha_2$ and $\alpha_1$ bands are visible.

To understand the effect of excess Fe on $\varepsilon_F$ (the unoccupied bandwidth of the $\alpha_2$ band) we analyze MDCs away from the low-energy region with the SC gap. The MDC peak positions from - 5 to - 50 meV are shown as red dots in Fig. 4 (D to F), with the three panels corresponding to the data in Fig. 1 (A to C respectively). We note that the MDC peaks are the same, within our resolution, below and above $T_c$. Parabolic fits to the MDC peak dispersion are shown as red curves in Fig. 4 (D,E,F) and lead to $\varepsilon_F$ estimates of 19 , 10 and 6 meV for the three samples in order of decreasing $y$. The uncertainty in $\varepsilon_F$ obtained from the fit (95 % confidence bounds) was lower than 1.5 meV for all samples.

We determine the SC energy gap from the EDC peaks in Fig. 4 (A to C), which correspond to the coherent Bogoliubov quasiparticles in the spectral function below $T_c$. The dispersion of the EDC coherence peaks is shown by blue dots in Fig. 4 (D,-F). Let us denote the experimentally determined Bogoliubov dispersion by $E(\mathbf{k})$. The SC gap $\Delta$ is given by the minimum of $|E(\mathbf{k})|$ along a $\mathbf{k}$-space cut perpendicular to the normal state Fermi surface, (here a radial cut through $\Gamma$).

The SC gap for all three samples is $\Delta = 3 \pm 0.5$ meV, in agreement with Ref. [15, 30]. We thus find $\Delta/\varepsilon_F$ = 0.16 (large $y$), 0.3 (small $y$) and 0.5 (smallest $y$). Thus the important dimensionless ratio $\Delta/\varepsilon_F$ which characterizes the pairing strength, increases monotonically with decreasing excess Fe, going from the BCS regime at large $y$ to the BEC regime at small $y$.

Another important evidence for the BCS-BEC crossover comes from the characteristic change in the coherence peak dispersion. In the BCS regime, we expect a bending back of $E(\mathbf{k})$ for $k \approx k_F$, whereas in the BEC regime, the dispersion is qualitatively different with a minimum gap at $\mathbf{k}=0$ (see Supplementary Materials). This evolution is evident in both the raw ARPES data of Fig. 1 and the EDC peak dispersion in Fig. 4 (blue dots).

We can understand this better from the Bogoliubov dispersion $E(\mathbf{k}) = \sqrt{(\varepsilon_k - \tilde{\mu})^2 + \Delta^2}$, where $\varepsilon_k = \hbar^2 k^2 / 2m^*$, $\tilde{\mu}$ is a renormalized chemical potential and $\Delta$ is the SC gap, assumed to be $k$-independent for simplicity. Exact numerical calculations [34] have shown that this form of the dispersion, well known from the mean-field theory, is accurate across the BCS-BEC crossover, provided one allows for renormalization of $m^*$ and $\tilde{\mu}$ due to interaction effects. In the weak coupling BCS limit, $\tilde{\mu} = \varepsilon_F$, and the dispersion shows back-bending at $k_F$. More generally, in the BCS regime where $\tilde{\mu} > 0$, $E_k$ exhibits back-bending at $k^* = (2m^* \tilde{\mu})^{1/2}$. As one goes through the crossover, $k^*$ decreases, which can also be described as the shrinking of the "minimum gap locus" [11]. In the BEC regime, when $\tilde{\mu} < 0$, the locus has shrunk to a single point, so that the dispersion has a minimum gap at $\mathbf{k}=0$.

We use the BCS-inspired spectral function $A(k,\omega) = u_k^2 \delta(\omega - E_k) + v_k^2 \delta(\omega + E_k)$, with broadened delta functions, as a simple way to model our ARPES data (see the Supplementary Materials for details). In Fig. 1 (D to F) we plot $A(k,\omega)$ without the Fermi function that cuts off the ARPES intensity for unoccupied states at positive energies. We choose $\varepsilon_F = 20$, 10 and 2.5 meV in panels (D,E and F) respectively, with a fixed $\Delta = 5$ meV. We adjust the chemical potential using $\tilde{\mu} = \varepsilon_F - \Delta^2 / 4\varepsilon_F$, the mean-field result for the two-dimensional BCS-BEC crossover [35]. We see from Fig. 1 (D-F) that this simple model captures the evolution of SC state dispersion in the ARPES data. This analysis also gives insight into the ARPES spectral weight, which is controlled by the momentum distribution $v_k^2$. On the BCS side of the crossover low-energy coherence peaks have significant spectral weight only near $k \approx k_F$, while on the BEC side there is significant spectral weight in large momentum range around k=0. The fact that we focus on a hole band, as opposed to an electron band, makes it much easier to observe the BEC-like regime using ARPES (see the Supplementary Materials).

**Discussion**

Our main results are summarized in the introduction and we conclude with a discussion of the interesting open questions about the BCS-BEC crossover in multiband superconductors raised by our work. In contrast to the extensive theoretical literature [5, 6, 7] on the BCS-BEC crossover in single band systems, motivated in large part by ultracold atom experiments, the crossover theory for multiband systems, especially nearly compensated semimetals, is much less developed; see, however, refs. [1,12, 14].

We note that the evolution of the minimum gap locus from a contour in the BCS regime to a point in $\mathbf{k}$ space is a general consequence of the BCS-BEC crossover even in a multiband system [12]. This also points to the interesting possibility that the crossover can be band-selective, namely the pairing could be in a BEC regime on one band while in the BCS regime on another. This could well be the case for $Fe_{1+y}Se_{0.4}Te_{0.6}$, where we focused only on the crossover in the $\alpha_2$ band. We found that as the Fermi energy of the $\alpha_2$ hole band near $\Gamma$ is reduced, that on the electron band near $M$ is increased. However, in the polarization geometry used, we do not see any SC state coherence peaks on the electron band, and new experiments are needed to see how the SC gap on that band changes with excess Fe.

The SC state BCS-BEC crossover reported in this paper raises several vital questions about the normal state. Do preformed pairs exist above Tc in Fe(Se,Te)? Is there a pairing pseudogap above Tc? We conclude with a discussion of the significance and implications of these issues.

A normal state pseudogap was first observed in the underdoped high Tc superconductors, with ARPES

playing a key role. We now understand that the cuprate pseudogap [36] has many facets, including the proximity to the Mott insulator, short range singlet correlations, charge density wave fluctuations and preformed pairing. In contrast, the pseudogap in the single-band BCS-BEC crossover originates from pairing. The normal state thus evolves from a Fermi liquid in the BCS limit to a normal Bose liquid in the BEC limit via the appearance of a pairing pseudogap in the crossover regime, as predicted early on [37]. This pseudogap has been observed in the ultracold Fermi gases using a spectroscopic technique [4] analogous to ARPES.

For a multiband superconductor, such as Fe(Se,Te), the question is more complex than in the one-band case. Although the gaps, Fermi energies, and dispersions can be examined separately on each band by ARPES, there are fundamental properties of superconductors, such as the transition temperature Tc and the superfluid density $\rho_s$, that depend on all the bands and their mutual coherence. An understanding of how $\rho_s$ and $T_c$ evolve with Fe doping would shed light on the important questions of SC fluctuations and pseudogap above Tc. We note that recent diamagnetism experiments [17] on FeSe provide evidence for a large SC fluctuation regime above Tc. To date, however, there is no spectroscopic evidence for, or against, a pseudogap above Tc. This key question deserves further investigation.

**Materials and Methods**

High quality single crystals of $Fe_{1+y}Se_xTe_{1-x}$ were grown using the modified Bridgman method. The stoichiometric amounts of high purity Fe, Se and Te powders were grinded, mixed and sealed in a fused silica ampoule. The ampoule was evacuated to a vacuum better than $10^{-5}$ torr and the mixture was reacted at 750°C for 72 hours. The resulting sinter was then regrinded and put in a double wall ampoule that was again evacuated to a vacuum better than $10^{-5}$ torr.

The ampoule was placed in a two-zone furnace with a gradient of 5°C/cm and slowly cooled from 1040° to 600° C at rate of 2° C/hour followed by a faster cooldown to 360°C for 24 hours. The resulting boule contained single crystals that could be separated mechanically. To change the amount of excess Fe we annealed the crystals for 48 hours in ampoules that were evacuated and then filled them with different pressures of oxygen.

Transport measurements were performed using a homebuilt setup based on an Oxford Teslatron-system. The resistivity and Hall resistance were measured using the van der Pauw method between room temperature and 2 K.

High resolution ARPES measurements were performed at the UE112_PGM-2b-1^3 beamline at BESSY, Berlin, Germany and at the CASSIOPÉE Beamline at SOLEIL (Saint Aubin, France) using a photon energy of 22 eV. The samples were cleaved in vacuum better than $5\times10^{-11}$ Torr at base temperature and measured for not more than 6 hours. The base temperature at Bessy was 1 K and at Soleil 7 K. The energy resolution was 4 meV in both beamlines.

**Acknowledgements**
We acknowledge support from the US-Israel BSF Grant No. 2014077. This work was supported in part by the Israeli Science Foundation under grant number 885/13. We thank HZB for the allocation of synchrotron radiation beamtime. We acknowledge SOLEIL for provision of synchrotron radiation facilities. The research leading to these results has received funding from the European Community's Seventh Framework Programme (FP7/2007-2013) under grant agreement n.312284.
MR was supported by the NSF Grant DMR-1410364. We thank the Aspen Center for Physics, supported by NSF grant PHY-1066293, for the hospitality during the writing of this paper.
**Author contributions**: S.R., A.R. and A.K. performed the ARPES measurements with the help of E.D.L.R, A.T.I, P.L.F and F.B. K.B.C performed the crystal growth. S.R. performed the transport measurements. S.R., M.R. and A.K. wrote the paper. M.R. and A.K. were responsible for overall project planning and direction.
**Competing interest:** The authors declare that they have no competing interests.
**Data and materials availability:** All data needed to evaluate the conclusions in the paper are present in the paper and/or the Supplementary Materials. Additional data related to this paper may be requested from the authors at amitk@physics.technion.ac.il

# Supplementary Materials

**Comparison between superconducting and normal state ARPES data**

In Fig. S1 we compare data for the $\varepsilon_F = 10\,\text{meV}$ sample with $T_c = 14\,\text{K}$ in the normal and superconducting states. In panels (A) and (B) we show raw data at 1K and at 22K. The data is normalized by the intensity above the Fermi-level produced by the second order light. The asymmetric intensity of the bands on positive and negative momenta at 22K is due to the circular polarization used at high temperatures. The dispersion of $\alpha_2$ was extracted from both data sets. We find the maxima in the MDC peaks using a gaussian model. The band dispersion and $\varepsilon_F$ were found to be identical in the normal and SC states within the experimental error bars. The dashed line in panels a and b represents the dispersion.

In panels c and d we show three EDC taken at the $\Gamma$ point and at $k = 0.1\,\text{Å}^{-1}$ and $k = 0.15\,\text{Å}^{-1}$, at the same two temperatures.

At low temperatures the coherence peaks can be seen in all three EDC while above $T_c$ there are no coherence peaks. The $\alpha_1$ band, although not crossing the Fermi-level, contributes to the spectral weight up to the Fermi-level. The EDC in Fig. 4 of the main text show clearly that the top of the $\alpha_1$ band is about 25meV below the Fermi-level. But due to the finite life-time of the quasiparticles associated with $\alpha_1$ this band does contribute some spectral weight at the $\Gamma$ point up to the Fermi-level, as can be seen in panel d. Below $T_c$ a sharp coherence peak emerges on top of the normal state EDC.

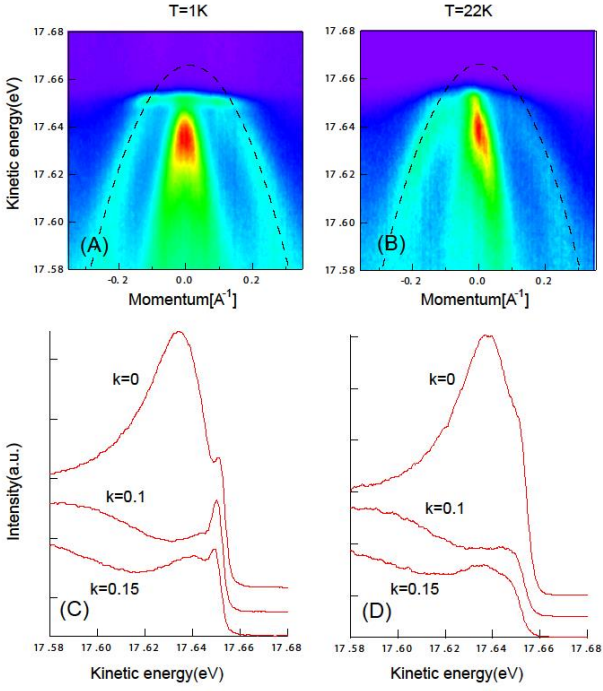

**Fig.S1 ARPES spectra above and below Tc** (A and B) ARPES spectra of a $\varepsilon_F = 10\,\text{meV}$ sample at T=1K and T=22K. The dashed line is the same parabolic dispersion from Fig. 4e in the main text. (C and D) Three EDCs at $k=0$, $k=0.1\,\text{Å}^{-1}$ and $k=0.15\,\text{Å}^{-1}$ at the same temperatures.

**Modeling the spectral function**

In Fig 1 (D-F) in the main text we plotted the spectral function $A(k,\omega)$ based on a simple model of the BCS-BEC crossover and it captures the qualitative features -- both dispersion and spectral weight -- of our ARPES data. ARPES probes only the occupied part of this spectral function, since it measures **k**-dependent matrix elements multiplying $f(\omega)A(k,\omega)$ convolved with resolution functions, where $f(\omega)$ is the Fermi function.

Here we carefully explain our modeling procedure focusing on the following points that we did not discuss in the main text.
(1) Line width broadening, (2) the renormalization of the chemical potential across the BCS-BEC crossover, and (3) the crucial negative signs that are important in modeling a hole band. We pay careful attention to the last point since most theory papers on the BCS-BEC crossover deal with "electron bands" with positive curvature.

We use a BCS-inspired form for the Green's function $G(k,\omega)$ and include a phenomenological line width broadening $\Gamma$ in the self energy [38] with

$$G^{-1}(k,\omega) = \omega - \xi_k + i\Gamma - \frac{\Delta^2}{\omega + \xi_k + i\Gamma} \quad (S1)$$

Here $\xi_k = \varepsilon_k - \mu$ is the dispersion $\varepsilon_k$ measured with respect to $\mu$, the chemical potential, $\omega$ is the energy, $\Delta$ the pairing gap and $\Gamma$ the broadening (taken to be independent of $\omega$ for simplicity).

The spectral function $A(k,\omega) = -\text{Im}\,G(k,\omega)/\pi$ is given by

$$A(k,\omega) = \frac{v_k^2}{\pi}\frac{\Gamma}{(\omega - E_k)^2 + \Gamma^2} + \frac{v_k^2}{\pi}\frac{\Gamma}{(\omega + E_k)^2 + \Gamma^2} \quad (S2)$$

Where $1 - u_k^2 = v_k^2 = \frac{1}{2}(1 - \xi_k/E_k)$ and $E_k = \sqrt{\xi_k^2 + \Delta^2}$. ARPES measurements will be dominated by the second term which is centered at $\omega = -E_k$ on the occupied side.

To model the $\alpha_2$ hole band in our ARPES data, we use

$$\varepsilon_k = -\frac{k^2}{2m^*} \text{ with } m^* = 3.2 m_e \quad (S3)$$

We also use a phenomenological $\Gamma = 3\,\text{meV}$.

It is crucial to take into account the renormalization of the chemical potential $\mu$ as one goes through the BCS-BEC crossover. Note that in our simple model we do not make a distinction between $\mu$ and $\tilde{\mu}$, where $\tilde{\mu}$ takes into account *additional* renormalization, beyond that which enters the thermodynamic $\mu$ specific to the excitation spectrum [6,34,39].

We next use the results of the mean field theory of the BCS-BEC crossover in 2D which yield analytical results [3] $\Delta = \sqrt{2\varepsilon_F E_A}$ and $\mu = \varepsilon_F - E_A/2$ that are valid across the entire crossover at $T=0$. (The $\mu$ result quoted here is for an electron-like band; see below).

Here $E_A$ is the binding energy of the two-body bound state in vacuum with $E_A/\varepsilon_F \ll 1$ corresponding to the weak-coupling BCS limit and $E_A/\varepsilon_F \gg 1$ to the extreme BEC limit.

Next we see, via a particle-hole transformation, that for a hole-band, with a negative curvature as in eq. S3, we must set $\mu$ to $-\mu$ in the equation above. We then eliminate (the experimentally unknown quantity) $E_A$ between the two equations to obtain

$$\mu = -\varepsilon_F + \Delta^2/4\varepsilon_F \qquad (S4)$$

It is easy to see that this is the correct sign convention, since in the noninteracting limit ($\Delta = 0$), the dispersion goes to the correct limit $\xi_k = (\varepsilon_k - \mu) \to -k^2/2m^* + \varepsilon_F$.

Finally, let us comment on the minimum gap locus and the spectral weights of the two Bogolyubov quasiparticle peaks as one goes through the BCS to BEC crossover. In the BCS regime, which for the hole band corresponds to $\mu < 0$ (note sign!), the minimum of the spectral gap $E_k = \Delta$ occurs at $\xi_k = 0$ corresponding to $k^* = \sqrt{2m^*|\mu|}$.

In this case the spectral weights $v_{k^*}^2 = u_{k^*}^2 = 1/2$. This is, of course, very well known in the extreme BCS limit with $\Delta \ll \varepsilon_F$, where $k^* \simeq k_F$.

In the BEC regime when $\mu > 0$ for a hole band, the minimum spectral gap $E_k = \sqrt{\mu^2 + \Delta^2}$ occurs at $\mathbf{k} = 0$. This is the collapse of the minimum gap locus discussed in the main text. The spectral weights at the gap edge $\mathbf{k} = 0$ are given by

$$v_{k=0}^2 = \frac{1}{2}\left(1 + \mu/\sqrt{\mu^2 + \Delta^2}\right)$$ which is (much) greater

than $u_{k=0}^2 = \frac{1}{2}\left(1 - \mu/\sqrt{\mu^2 + \Delta^2}\right)$.

It is instructive to look at the extreme BEC limit $\mu \gg \Delta \gg \varepsilon_F$, where $v_{k=0}^2 \simeq 1$ while $u_{k=0}^2 \simeq \mu^2/4\Delta^2 \ll 1$

Thus there is a large spectral weight $v_k^2$ on the occupied side that can be measured by ARPES for a hole-band in the BEC regime. In contrast, the roles of $u_k^2$ and $v_k^2$ are interchanged for an electron band, and one would have a tiny occupied spectral weight in the BEC regime. In this sense, the BCS-BEC crossover in the hole band investigated in this paper is ideally suited for being probed by ARPES.

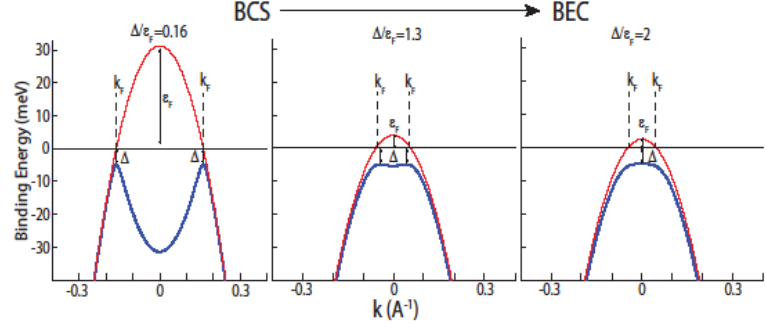

**Fig. S2 Bogoliubov dispersion from BCS to BEC.** Evolution of the Bogoliubov dispersion from BCS state, where the bending-back at kF can be seen to the BEC state where the minimum of the gap is found at k=0.